\documentclass[prl,a4paper,reprint,superscriptaddress,nofootinbib,preprintnumbers]{revtex4-1}
\pdfoutput=1

\usepackage{mymacros}

\begin{document}

\preprint{YITP-18-80}

\title{Complexity as a novel probe of quantum quenches: universal scalings and purifications}

\author{Hugo A. Camargo} \email{hugo.camargo@aei.mpg.de}
\affiliation{Max  Planck  Institute  for  Gravitational  Physics (Albert Einstein Institute),\\Am M\"uhlenberg 1, 14476 Potsdam-Golm, Germany}
\affiliation{Department of Physics, Freie Universit\"{a}t Berlin, Arnimallee 14, 14195 Berlin, Germany}

\author{Pawel Caputa} \email{pawel.caputa@yukawa-kyoto-u.ac.jp}
\affiliation{Center for Gravitational Physics, Yukawa Institute for Theoretical Physics (YITP),\\Kyoto University, Kitashirakawa Oiwakecho, Sakyo-ku, Kyoto 606-8502, Japan.}

\author{Diptarka Das} \email{diptarka.das@aei.mpg.de}
\affiliation{Max  Planck  Institute  for  Gravitational  Physics (Albert Einstein Institute),\\Am M\"uhlenberg 1, 14476 Potsdam-Golm, Germany}

\author{Michal P.\ Heller} \email{michal.p.heller@aei.mpg.de}
\altaffiliation[\emph{On leave of absence from:}]{ National Centre for Nuclear Research, Ho{\.z}a 69, 00-681 Warsaw, Poland.}
\affiliation{Max  Planck  Institute  for  Gravitational  Physics (Albert Einstein Institute),\\Am M\"uhlenberg 1, 14476 Potsdam-Golm, Germany}

\author{Ro Jefferson} \email{rjefferson@aei.mpg.de}
\affiliation{Max  Planck  Institute  for  Gravitational  Physics (Albert Einstein Institute),\\Am M\"uhlenberg 1, 14476 Potsdam-Golm, Germany}


\begin{abstract}
We apply the recently developed notion of complexity for field theory to a quantum quench through a critical point in $1\!+\!1$ dimensions. We begin with a toy model consisting of a quantum harmonic oscillator, and show that complexity exhibits universal scalings in both the slow and fast quench regimes. We then generalize our results to a 1-dimensional harmonic chain, and show that preservation of these scaling behaviours in free field theory depends on the choice of norm. Applying our set-up to the case of two oscillators, we quantify the complexity of purification associated to a subregion, and demonstrate that complexity is capable of probing features to which the entanglement entropy is insensitive. We find that the complexity of subregions is subadditive, and comment on potential implications for holography. 
\end{abstract}

\maketitle

\begin{center}\textbf{INTRODUCTION}\end{center}

Among the most exciting developments in theoretical physics is the confluence of ideas from quantum many-body systems, quantum information theory, and gravitational physics. Recent progress in this vein includes the development of tensor network methods for simulating quantum many-body systems (see, e.g., \cite{Orus:2013kga}), proofs of irreversibility of RG flows using quantum information techniques~\cite{Casini:2012ei,Casini:2004bw,Myers:2010xs,Myers:2010tj,Casini:2015woa,Casini:2017vbe}, and the illumination of the role of codimension-2 extremal surfaces in the emergence of holographic spacetime (see, e.g., \cite{Rangamani:2016dms}). The central technical tool in these ground-breaking results is the reduced density matrix for a spatial subregion, and the associated von Neumann entropy, cf. \cite{Ryu:2006bv,Hubeny:2007xt}. 

However, insights from black hole physics~\cite{Susskind:2014rva,Susskind:2014moa,Brown:2015bva,Brown:2015lvg} suggest that certain codimension-0 and codimension-1 surfaces may also play an important role in reconstructing bulk spacetime in holography, since these capture information beyond that which is accessible to the aforementioned codimension-2 surfaces---that is, beyond entanglement entropy. These geometric objects are conjectured to be dual to the ``complexity'' of the boundary field theory, according to the competing ``complexity=volume'' (CV)~\cite{Susskind:2014rva,Susskind:2014moa} and ``complexity=action'' (CA) proposals~\cite{Brown:2015bva,Brown:2015lvg}.

Drawing on earlier developments~\cite{Nielsen:2005mn1,Nielsen:2006mn2,Nielsen:2007mn3,Haegeman:2011uy,Nozaki:2012zj}, \cite{Jefferson:2017sdb} and \cite{Chapman:2017rqy} sought to make the above conjectures more precise by defining the notion of complexity in (free, bosonic) quantum field theory (this idea was subsequently extended to fermionic theories in~\cite{Hackl:2018ptj}, see also~\cite{Khan:2018rzm,Reynolds:2017jfs}; for alternative approaches to defining complexity in field theories, see~\cite{Caputa:2017urj,Caputa:2017yrh,Czech:2017ryf,Bhattacharyya:2018wym,Magan:2018nmu,Caputa:2018kdj,Yang:2018nda}). In light of the successes born of entanglement entropy mentioned above, understanding complexity in quantum field theory represents a very promising research direction. Particularly interesting open questions include the time-dependence of complexity, and the interplay between complexity and entanglement entropy in non-equilibrium systems. It is therefore of value to have a tractable system in which these ideas can be concretely explored. 

To that end, one of the most active areas of research into non-equilibrium quantum dynamics is the study of quantum quenches \cite{Gring1318, Polkovnikov:2010yn}, in which remarkable progress has been made in understanding the mechanisms underlying thermalization encoded in the reduced density matrix~\cite{Dunjko}. Theoretical studies within the scope of experimental verification have revealed that smooth quenches through a critical point exhibit universal signatures via scalings. The Kibble-Zurek (KZ) scaling~\cite{Kibble:1976sj,Zurek:1985qw} is the most well-known example of this behaviour, and has received a great deal of attention in recent years \cite{Chandran_etal, Mondal_etal,Gritsev:2009wt,Dziarmaga,Lamacraft}. In this case, the state is evolved adiabatically until very close to the critical point, and hence the regime of KZ can be characterized as ``slow''. Recent studies in holography~\cite{Buchel:2012gw,Buchel:2013lla}, free field theory~\cite{Buchel:2013gba,Das:2014jna,Das:2014hqa}, and lattice spin models~\cite{Das:2017sgp} have also revealed new scaling behaviours in a ``fast'' (non-adiabatic) regime. This fast scaling behaviour appears to be a universal feature of any interacting theory which flows from a CFT in the ultraviolet (UV)~\cite{Das:2016lla,Goykhman:2018ihr,Dymarsky:2017awt}. At a technical level, previous studies have mainly focused on the scalings of a restricted set of one- and two-point functions, and recently on entanglement~\cite{Caputa:2017ixa}. However, as we shall argue below, the latter probes at most only a spatial subsystem, while complexity is a property of the entire wavefunction. Hence complexity represents a means of probing features of quench dynamics to which entanglement entropy is insensitive. Initial steps towards applying complexity to quenches were taken in \cite{Alves:2018qfv}, for a quench which monotonically interpolates between two massive theories.

Motivated by these scaling phenomena, we explore the complexity of exact \emph{critical} quench solutions for free scalar theories, and find evidence for universal scaling behaviour. Our primary model will consist of a bosonic oscillator whose frequency varies smoothly with time, and asymptotes to a finite constant in both the far future and past. We first define complexity for a single mode, and then generalize our results to a 1-dimensional harmonic chain. However, we find that a judicious definition of complexity is required in order to make the scaling expectations for free field theory manifest. Utilizing this set-up, we contrast the complexity and the entanglement entropy for a fixed bipartition of the Hilbert space of two coupled harmonic oscillators. This model enables us to quantify the notion of ``complexity of purification'' recently introduced in \cite{Agon:2018zso}, which allows one to associate a complexity to subregions (i.e., mixed states). We find that the complexity of subregions is subadditive, which may have interesting implications for the CV vs. CA proposals above.

\begin{center}\textbf{COMPLEXITY OF QUANTUM QUENCHES}\end{center}

\begin{center}\textbf{Quench model}\end{center}

We shall begin with the following simple Hamiltonian describing a free bosonic oscillator:
\be
H(t) = \frac{1}{2M}P^2 + \frac{1}{2}M\tilde\omega^2 X^2~,\label{eq:Hphys}
\ee
where $M$ is the mass of the oscillator, $\tilde\omega(t/\delta t)$ is some time-dependent frequency profile with an intrinsic scale set by the parameter $\delta t$, and the canonical position and momentum operators satisfy $[X,P]=i$. However, for reasons that will become apparent below, it is preferable to work with the dimensionless variables $x\!\equiv\!\varpi X$, $p\!\equiv\! P/\varpi$, $\omega\!\equiv\!\tilde\omega/\varpi$, where $\varpi$ is some new mass scale, which will be given an interpretation as the gate scale when we introduce our quantum circuit (see appendix A). Setting $\varpi=M$ for simplicity, \eqref{eq:Hphys} becomes
\be
H(t) = M\lp\frac{p^2}{2} + \omega^{2}\frac{x^2}{2}\rp, \label{eq:H}
\ee
where the quantities appearing in the parentheses are all dimensionless, and we shall henceforth set $M=1$. The time-evolved initial ground-state wavefunction at time $t$ for the Hamiltonian \eqref{eq:H} takes the form
\be
 \psi_0(x,t)=\mathcal{N}\exp\left(\frac{i}{2}\frac{\dot{f}^*}{f^*}x^2\right),\label{eq:ground}
\ee
where $\mathcal{N}\equiv\lp2\pi f^*f\rp^{-1/4}$, and $f(t/\delta t)$ is the solution to the equation
\be
\ddot{f}+\omega^2 f=0~.\label{eq:eqf}
\ee
Now, we desire a quench profile $\omega^2(t/\delta t)$ which admits an exact solution to this equation, and which asymptotes to a constant at both early and late times, with changes occurring in the time-window $[-\delta t,\delta t]$. One of the most common profile used in the literature (see, e.g., \cite{Caputa:2017ixa}) is 
\be
\omega^{2}(t/\delta t) = \omega_{0}^2 \left( 1-\frac{1}{\cosh^2\!\left( \frac{t}{\delta t}\right) } \right)~.\label{eq:profile}
\ee
Here $\omega_{0}$ is a free parameter, but will gain an interpretation as the dimensionless reference-state frequency below. This profile has the property that the system is initially gapped at $t=-\infty$, but becomes gapless at $t=0$, corresponding to oscillator excitations above the ground state \eqref{eq:ground} as the system evolves via \eqref{eq:H}. In this case, the function $f(t)$ can be written explicitly in terms of hypergeometric functions---see \cite{Caputa:2017ixa}.

Our interest in this set-up is due to the fact that it can also be used to study the ground state of two (or more) harmonic oscillators with a time-dependent coupling. The same model was considered in \cite{Bombelli:1986rw,Srednicki:1993im,Caputa:2017ixa} for investigating entanglement entropy during a quench. Explicitly, the Hamiltonian for two oscillators is given by
\be
H=\frac{1}{2}\left[p^2_1+p^2_2+ 2\Omega^2  \left(x_1 - x_2\right)^2+\omega^2\left( x_1^{2} + x_2^{2} \right)\right]~.
\label{eq:H2}
\ee
In the normal-mode basis $x_\pm=(x_1\pm x_2)/\sqrt{2}$, this Hamiltonian takes the decoupled form
\be
H(t)=H_{+}(t)+H_{-}(t)\label{eq:2sho}
\ee
where the subscript denotes the use of the $\pm$ mode in \eqref{eq:H}, with $\omega^2_+ = \omega(t)^2$ and $\omega^2_- = \omega(t)^2 + 4 \Omega(t)^2$. The corresponding wavefunction is then given by
\be
\psi(x_{+},x_{-},t) =\psi_0(x_+,t)\psi_0(x_-,t)~,\label{eq:psiProd}
\ee
with $\psi_0$ given by \eqref{eq:ground}. Note that this construction naturally generalizes to an $N$-oscillator harmonic chain, which we will consider after introducing complexity below.

\begin{center}\textbf{Circuit complexity}\end{center}
To evaluate the complexity of the target state \eqref{eq:psiProd}, we shall apply the circuit complexity approach of \cite{Jefferson:2017sdb}, adapted at the level of covariance matrices as in \cite{TFD}. The reader is referred to these works for details. In brief, a circuit $U$ is a unitary operator whose action on some reference state $\psi_\mr{R}$ produces the desired target state $\psi_\mr{T}$,
\be
\ket{\psi_\mr{T}}=U\ket{\psi_\mr{R}}~.\label{eq:evo}
\ee
In analogy with quantum circuits, $U$ can be thought of as a sequence of fundamental gates, each of which effects an infinitesimal change to the state. The complexity of the target state is then defined as the length of the optimum circuit according to some suitably chosen depth function (e.g., the number of gates). Note the keyword ``optimal'': there may be arbitrarily many different circuits which satisfy \eqref{eq:evo}. Hence the central feature of \cite{Jefferson:2017sdb} was to use the geometric approach of Nielsen and collaborators \cite{Nielsen:2005mn1,Nielsen:2006mn2,Nielsen:2007mn3} to convert the problem of finding the optimum circuit into that of identifying the minimum geodesic in the geometry generated by the algebra of gates. 

Given the form of \eqref{eq:psiProd}, it is sufficient to begin with a single oscillator. Hence we are interested in target states of the form
\be
\psi_\mr{T}(x,t)=\lp\frac{a}{\pi}\rp^{1/4}\exp\left\{-\frac{1}{2}\lp a+ib\rp x^2\right\}~,\label{eq:psiT}
\ee
where $a(t),b(t)\in\reals$ are the real and imaginary parts of the frequency $i\dot{f}^*/f^*$ in \eqref{eq:ground}, and we have suppressed the time-dependence for compactness. Note that $a\!>\!0$ (one can verify that the solutions to \eqref{eq:eqf} indeed satisfy this normalizability constraint), while $b$ may take any sign. Our reference state will be provided by the ground state of our time-dependent Hamiltonian \eqref{eq:H2} at $t\!=\!-\infty$,
\be
\psi_\mr{R}(x)=\lp\frac{\omega_\mr{R}}{\pi}\rp^{1/4}\exp\left\{-\frac{\omega_\mr{R}}{2}x^2\right\}~,\label{eq:psiR}
\ee
where $0\!<\!\omega_\mr{R}\!\in\!\reals$. Our task is now to construct a circuit $U$ satisfying \eqref{eq:evo} according to the geometric approach outlined above. 

The details of our complexity calculation are given in appendix A. The key point is that we may view $U$ as a matrix which acts at the level of covariance matrices, so that \eqref{eq:evo} becomes
\be
G_\mr{T}=UG_\mr{R}U^T~,\label{eq:evoG}
\ee
where the matrix elements of $G$ are given by
\be
G^{ab}=\bra{\psi}\xi^a\xi^b+\xi^b\xi^a\ket{\psi}~,
\ee
where $\xi^a\!\equiv\!\{x^1,p^1,\ldots x^N,p^N\}$ are the dimensionless phase-space operators for $N$ oscillators. The covariance matrix is an equivalent representation of the wavefunction, which has the advantage of making the explicit choice of gates more transparent. In particular, we seek the minimal set of gates necessary to effect the desired transformation. As explained in appendix A, this naturally leads to hyperbolic space, with the metric
\be
\dd s^2=\frac{2\dd z^2+\dd y^2}{8z^2}~,
\ee
and therefore the complexity of the target state \eqref{eq:psiT} is given by the well-known geodesic distance formula on $\mathbb{H}^2$ (cf. appendix B), which admits a particularly compact expression in terms of the squeezed target-state covariance matrix $\tilde G_\mr{T}=SG_\mr{T}S^T$:
\be
\CC=\frac{1}{2}\ln\lp\chi+\sqrt{\chi^2-1}\rp~,\;\;\;\;\;
\chi\equiv\frac{1}{2}\mr{tr}\,\tilde G_\mr{T}~,\label{eq:C}
\ee
where $S$ is the squeezing operator defined such that $SG_\mr{R}S^T=\mathbbm{1}$. This result immediately generalizes to the case of $N$ oscillators: since $\tilde G_\mr{T}$ is block-diagonal in an appropriate basis, the geometry factorizes into $N$ independent copies of $\mathbb{H}^2$. Hence the complexity of a 1-dimensional lattice of oscillators is
\be
\CC=\sqrt{\sum_{j=1}^{N}\left[\frac{1}{2} \ln\lp\chi_{j}+\sqrt{\chi_{j}^2-1}\rp\right]^2}.\label{eq:Cfield}
\ee
Note that in this expression, we have added the complexities in the $L_2$-norm; we shall comment on the use of other norms in appendix D. By taking the continuum limit of such a lattice, we obtain the complexity for a bosonic system in $1\!+\!1$ dimensions. Specifically, we consider the harmonic chain whose Hamiltonian is given by
\be\label{eq:Hfield}
H= \frac{1}{2} \sum_{n=1}^{N} \left( \Pi_n^2 + ( \phi_{n+1} - \phi_n )^2 + m^2(t) \phi_n^2 \right)~,
\ee
where $(\phi_n, \Pi_n)$ are mutually conjugate scalar field variables. Since we work with dimensionless variables, we shall set the lattice spacing (i.e., the UV-cutoff) to unity. In momentum space, each mode then satisfies
\be
\ddot{\phi_k} + \left( 4 \sin^2 \frac{k}{2} + m^2(t) \right) \phi_k = 0~,
\label{eq:mode}
\ee
where we have imposed periodic boundary conditions $k\!=\!k\!+\!2\pi$, and the quench profile is given by $m(t)\!=\!\omega(t/\delta t)$ in \eqref{eq:profile}. The reference state, $\ket{\psi_R}$ is given by the ground state of the Hamiltonian \eqref{eq:Hfield} at $t=-\infty$ when $m(t) = \omega_0$. Integrating over momentum modes, the continuum limit of \eqref{eq:Cfield} is simply
\be
\CC(t)=\sqrt{\int_0^{2\pi} \frac{\dd k}{2\pi} \left[ \frac{1}{2} \ln\lp\chi_{k}(t)+\sqrt{\chi_{k}^2(t)-1}\rp\right]^2}.\label{eq:Cfield2}
\ee 
where $\chi_k(t)$ is given in \eqref{eq:C} with the covariance matrix corresponding to the $k^{\mathrm{th}}$ oscillator. 
\begin{figure}[h!]
\centering
\includegraphics[width=0.95\columnwidth]{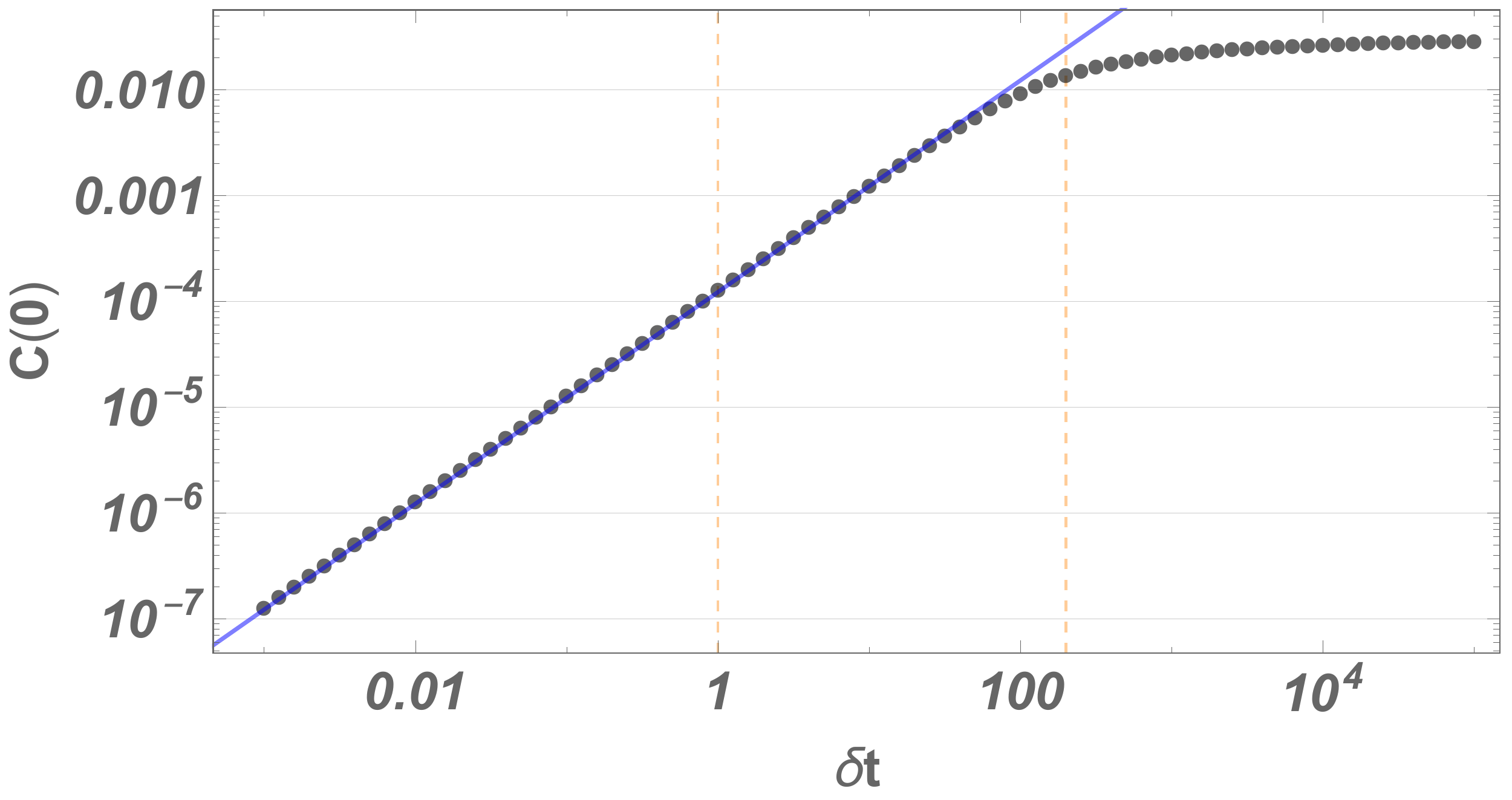}
\caption{Log-log plot of complexity of the $(1\!+\!1)$-dimensional free field theory \eqref{eq:Cfield2} at the critical point $t\!=\!0$ vs. the quench rate $\delta t$ (measured in units of the lattice spacing), with $\omega_0\!=\!0.005$. The straight-line fit (blue) reveals linear scaling in the fast regime.\label{fig:QC-field}}
\end{figure}

Since we are interested in the behaviour of complexity as the system passes through the critical point of the quench, it is sufficient to evaluate this function at $t=0$; see Fig.~\ref{fig:QC-field}. This then allows us to extract the universal scaling behaviours, which we examine in more detail in the next section.

\begin{center}\textbf{Universal scalings in complexity}\end{center}

\begin{figure}[h!]
\centering
\includegraphics[width=0.95\columnwidth]{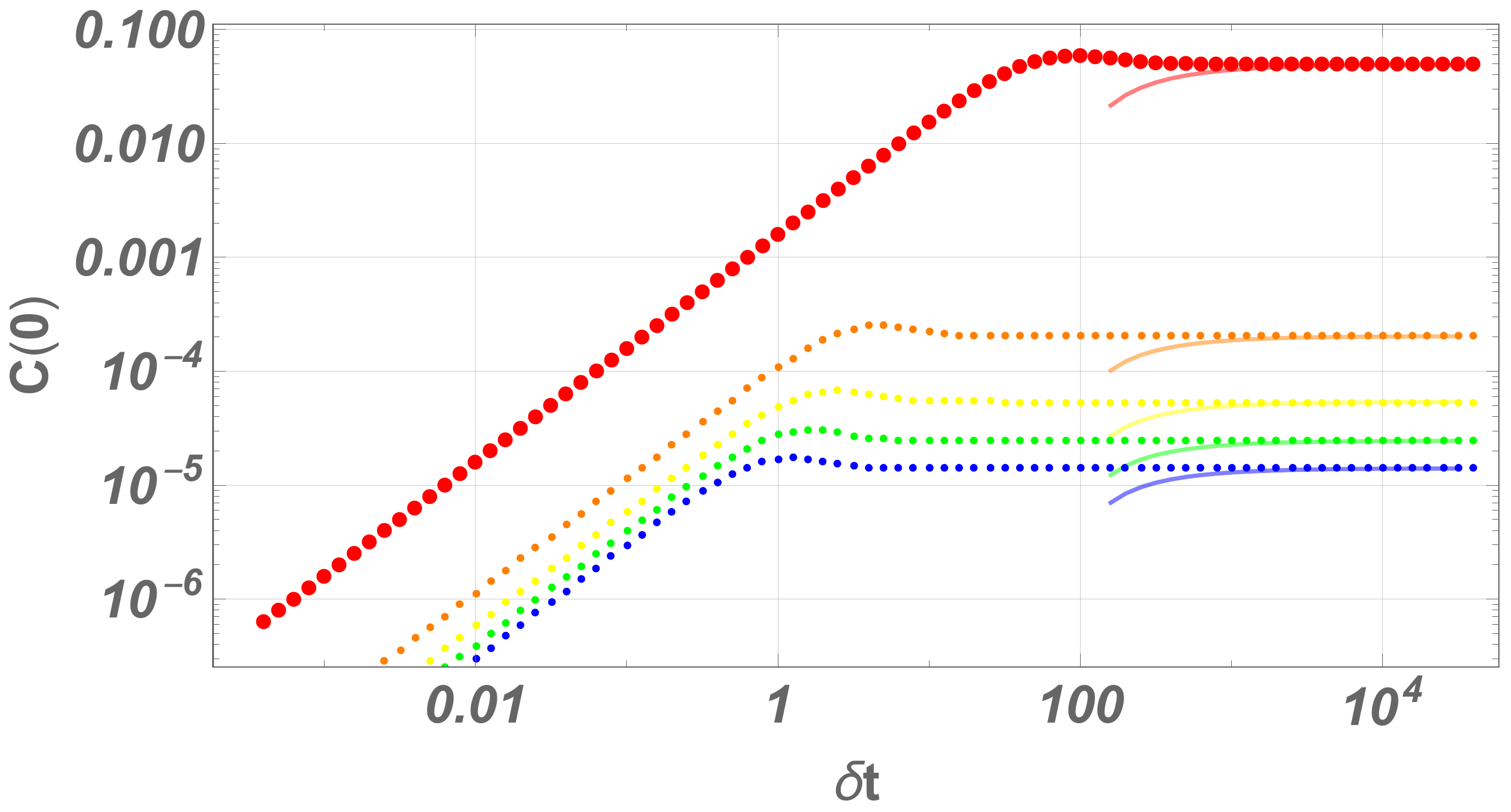}
  \caption{Single-mode contributions to the complexity \eqref{eq:Cfield2} at the critical point $t\!=\!0$ for $\omega_0\!=\!0.005$ and $k\!=\!\{0.006, 0.111, 0.216, 0.320, 0.425 \}$ (resp. red, orange, yellow, green, blue). For large $\delta t$, the exact solutions (dotted) agree with the saturation values \eqref{eq:Csat} predicted from KZ (solid).\label{fig:QC-mode}}
\end{figure}

We now wish to examine the presence of universal scalings of the critical complexity with respect to the quench rate. In particular, the contributions from individual momentum modes to $\CC(0)$ in \eqref{eq:Cfield2} are plotted in Fig.~\ref{fig:QC-mode}. We find that all modes go to zero in the sudden-quench limit $\delta t\rightarrow0$, which is consistent with results for instantaneous quenches. For all $k>0$, we observe mode-dependent saturation in the slow regime $\delta t\rightarrow\infty$, consistent with what one expects from KZ. In the adiabatic approximation, the KZ scale arises from the Landau criterion for the breakdown of adiabaticity,
\be
\frac{1}{E(t)^2 } \frac{\dd E(t)}{\dd t}\bigg\vert_{t_\mr{KZ} } =  1~, \label{eq:KZ-cond}
\ee
where $t_\mr{KZ}$ is the Kibble-Zurek time and $E$ is the time-dependent mass gap from criticality. For the profile \eqref{eq:profile}, one finds $t_\mr{KZ}\!\approx\!\sqrt{ \delta t/ \omega_0}$, at which time the frequency is
$\omega_\mr{KZ}(k)\!=\!\sqrt{ 4 \sin^2 \tfrac{k}{2} + m^2(t_\mr{KZ}) }
\!\approx\!\sqrt{ 4 \sin^2 \tfrac{k}{2} +  \tfrac{\omega_0}{\delta t} }$, where we have used the fact that $m^2(t_\mr{KZ})\!\sim\! \omega_0^2 t_\mr{KZ}^2/\delta t^2$, since in the slow regime $\delta t\!>\!t_\mr{KZ}$ by definition. Hence the KZ scaling for the $k^\mathrm{th}$ mode may be extracted by calculating the complexity at this frequency. One finds logarithmic KZ scaling in the slow regime for $\delta t\!<\!\frac{\omega_0}{4}\csc^2 \frac{k}{2}$. As soon as $\delta t$ exceeds this value, we observe saturation in the frequency (to $2 \sin \frac{k}{2}$), and hence also in complexity to
\be
\CC_{\mr{sat}}^k = \frac{1}{2} \log \left(  \frac{\sqrt{ \omega_0^2 + 2 - 2 \cos k}}{2 \left|\sin \tfrac{k}{2}\right|} \right) ~.\label{eq:Csat}
\ee
The KZ approximation is superimposed on the exact results in Fig.~\ref{fig:QC-mode}, which clearly shows agreement with the saturation value \eqref{eq:Csat} in the large-$\delta t$ limit.

\begin{figure}[h!]
\centering
\includegraphics[width=0.95\columnwidth]{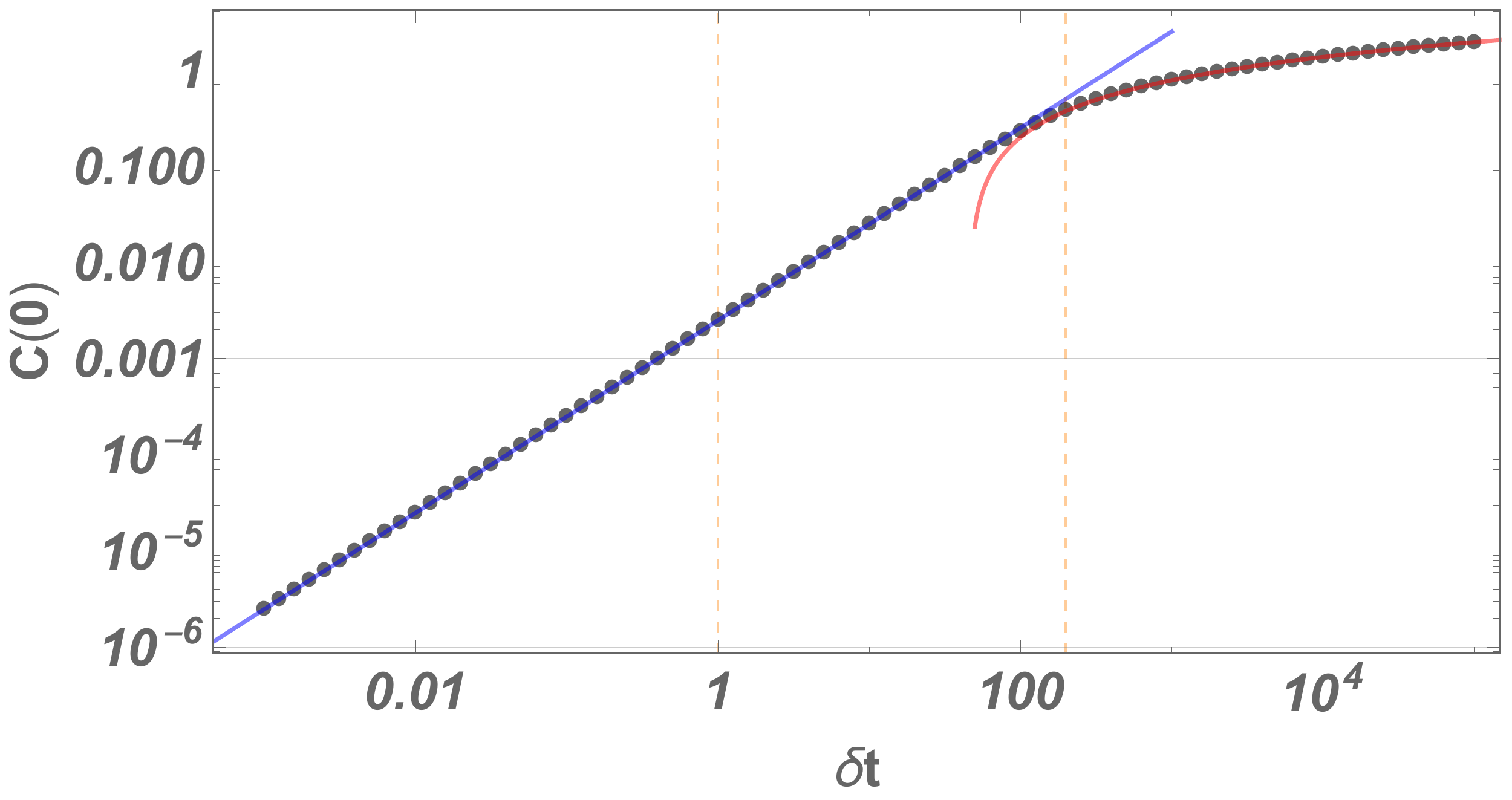}
\caption{Zero-mode contribution to $\CC(0)$ \eqref{eq:Cfield2} as a function of the quench rate $\omega_0\delta t$, with $\omega_0\!=\!0.005$. The complexity scales linearly in the fast regime ($\ln C/\ln \delta t=1$, blue), and smoothly transitions to a logarithmic scaling $\tfrac{1}{4}\log \delta t$ in the slow regime (red). The transition to KZ occurs at $\omega_0\delta t\!\sim\!1$ which in this case is $\delta t\!\sim\!200$ in lattice units.\label{fig:QC-scalings}}
\end{figure}

The critical complexity of the zero mode $k\!=\!0$ exhibits universal scalings in both the slow and fast regimes. Indeed, this same behaviour is exhibited by the single quantum oscillator we initially introduced upon sending the frequency to zero (i.e., we take the $\omega_-$ solution for the two-oscillator case above). Unlike higher modes, the zero mode does not saturate at large $\delta t$ since the logarithmic scaling is always present. From the KZ analysis above, we can derive the universal coefficient of the log as $\frac{1}{4}$, which is confirmed by fitting the exact solution, as shown in Fig.~\ref{fig:QC-scalings}. We note that the KZ scaling exhibited by entanglement entropy under a critical quantum quench has the same form, but with a $\frac{1}{6}$ coefficient instead \cite{Caputa:2017ixa, Nishida:2017hqd}. Meanwhile in the fast regime ($\delta t<1$ in lattice units), the complexity grows linearly with $\delta t$. While these scalings are present for higher modes as well, they are confined to increasingly narrow regions of $\delta t$ for larger values of $k$.

\begin{center}\textbf{COMPLEXITY VS. ENTANGLEMENT}\end{center}

One of the main motivations for the holographic complexity proposals was the observation that the information contained in the reduced density matrix of any spatial bipartition of the CFT Hilbert space, as encoded in the entanglement entropy, is generally insufficient to determine the entire bulk geometry \cite{Susskind:2014moa} (see also \cite{Freivogel:2014lja} and references therein). One can then ask whether complexity provides another take on the information contained in reduced density matrices. Indeed, recent proposals for the complexity of subregions in holography -- that is, on the bulk side -- have been made in \cite{Carmi:2016wjl,Alishahiha:2015rta,Ben-Ami:2016qex,Abt:2017pmf,Abt:2018ywl}. However, since the field-theoretic notion of complexity above is defined for pure states, it is not \emph{a priori} clear how to define complexity for the reduced density matrix corresponding to some spatial subregion. 

A particularly natural extension of existing pure-state definitions to this case is the \emph{complexity of purification}, recently outlined in \cite{Agon:2018zso}, in which the complexity of the subsystem is defined by minimizing over the complexities of all possible purifications (see also \cite{Stoltenberg:2018ink}). Applying our quench set-up above to the case of two oscillators allows us to quantify this proposal, by considering the reduced density matrix corresponding to a single oscillator, say $x_1$, and purifying within the original Hilbert space of Gaussian states (i.e., without ancilla). The total wavefunction depends on six real parameters, three of which we fix by our knowledge of the covariance matrix for oscillator $x_1$. Minimizing over the remaining three parameters then gives the complexity of purification for the subsystem, which we shall denote $\CC_A$ in reference to a generic subsystem $A$ and its complement $\bar{A}$. 
\begin{figure}[h!]	 
\includegraphics[width=0.95\columnwidth]{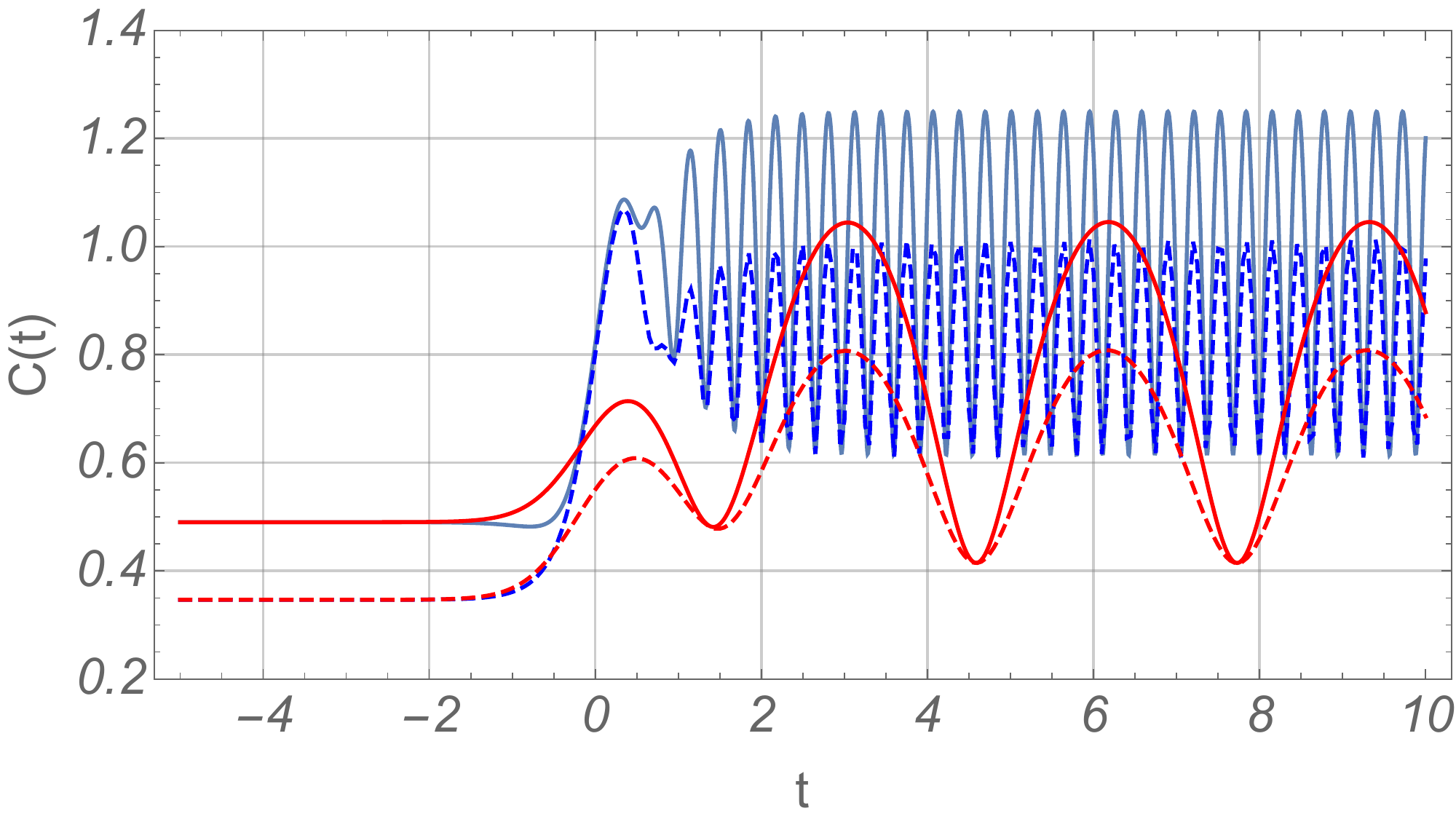}
\caption{Comparison of the complexity \eqref{eq:Cfield} as a function of time $t$ of the original target state (solid) and the optimum purification (dashed) for $\delta t=10$ (blue) and $\delta t=1$ (red), with $\omega_\mr{R}=0.5$ for both oscillators. Note that the latter never exceeds the former, and is always greater than $\CC/2$; that is, the complexity of purification appears to satisfy superadditivity \eqref{eq:super}. We have tested this conjecture numerically for $\sim\!70,\!000$ cases.\label{fig:purification}}
\end{figure}

As observed in Fig.~\ref{fig:purification}, the complexity of purification satisfies $\CC/2\leq\CC_A\leq\CC$, which we have verified numerically for a wide range of values in the six-parameter landscape spanned by the components of the covariance matrix. The upper inequality is saturated iff the original target state happens to be the least complex state among all possible purifications. Meanwhile, the lower inequality is saturated iff the original target state is a product state with respect to the chosen bipartition; i.e., subsystem $A$ describes a pure state, $S_A=0$. This can be understood from the fact that the purification process seeks to produce a state which is as close to the reference state as possible, since the latter has minimum complexity by fiat. Since in this case the reference state is an unentangled product state, the minimum purification is one in which the complement $\bar A$ is also an unentangled product state---but this is only possible if the original state is a tensor product of the form $\HH_A\otimes\HH_{\bar A}$, otherwise the entanglement across the bipartition prevents one from obtaining the reference state in the restriction $\bar A$. While one should exercise caution in blithely generalizing from this simple two-oscillator case, the above leads us to suggest that the complexity of subsystems is \emph{subadditive}:
\be
\CC_A+\CC_{\bar A}\geq\CC~.\label{eq:super}
\ee
As observed in \cite{Agon:2018zso}, this agrees with the holographic CA proposal, but not with the CV proposal, which is superadditive. \\

\begin{center}\textbf{OUTLOOK}\end{center}
Quenches represent tractable models of dynamical quantum systems in which complexity can be better understood, as well as yield new physical insights; e.g., we have found that complexity can be used to extract universal scalings. We have also examined the complexity of subregions (i.e., mixed states) via their purifications. Since complexity encodes global information about the state, it is sensitive to features to which entanglement is blind. We find that subregion complexity appears to satisfy superadditivity \eqref{eq:super}, which is consistent with the CA proposal. While it would be premature to take definitive lessons for holography from such simple free field models, this may provide further hints as to the proper notion of complexity in holographic field theories, and thereby shed light on ongoing efforts to reconstruct bulk spacetime in AdS/CFT.

\begin{center}\textbf{ACKNOWLEDGMENTS}\end{center}
We thank E.~M.~Brehm, S.~R.~Das, D.~A.~Galante, E.~L\'opez, J.~Magan, T.~Takayanagi, M.~Walter, and the co-authors of \cite{TFD} for helpful conversations. We would especially like to thank H.~Marrochio, R.~C.~Myers, and M.~Smolkin for comments on a draft of this manuscript. HC, DD, MH and RJ acknowledge the hospitality of MITP during the ``Modern Techniques for AdS and CFT'' program during the completion of this project. HC is partially supported by the Konrad-Adenauer-Stiftung through their Sponsorship Program for Foreign Students. PC is supported by the Simons Foundation through the ``It from Qubit'' collaboration and by the JSPS starting grant KAKENHI 17H06787. The Gravity, Quantum Fields and Information group at AEI is generously supported by the Alexander von Humboldt Foundation and the Federal Ministry for Education and Research through the Sofja Kovalevskaja Award.

\bibliographystyle{apsrev4-1}
\bibliography{biblio}

\clearpage

\appendix{}

\begin{center}\textbf{SUPPLEMENTARY MATERIAL (APPENDICES)}\end{center}

\section{A. Circuit complexity for $\mathbb{H}^2$}\label{appx:complexity}

In this appendix, we explain the calculation of the complexity of the state \eqref{eq:psiT}. Our approach closely follows that of \cite{Jefferson:2017sdb}, but adapted at the level of covariance matrices as in \cite{TFD}; the reader is referred to these works for more details. As stated in the main text, one seeks the optimum circuit $U$ which acts on the reference state to produce the target state according to \eqref{eq:evo}. Identifying, and associating a well-defined length to, this optimum circuit requires geometrizing the problem \cite{Nielsen:2005mn1,Nielsen:2006mn2,Nielsen:2007mn3}. To proceed, one represents the circuit as a path-ordered exponential,
\be
U(s)=\cev{\PP}\exp{\int_0^s\!\dd\tilde s\,Y^I\!(\tilde s)M_I}~,\label{eq:Um}
\ee
where the path parameter $s$ is chosen to run from 0 at the reference state to 1 at the target state (i.e., $s\!=\!1$ in \eqref{eq:evo}). The matrix generators $M_I$ represent the algebra of gates, while the parameters $Y^I$ can the thought of as turning these gates on and off at specified points along the path (these will be given a precise interpretation below as the components of the frame bundle that we use to construct the inner product, and thereby the geometry). However, in order to obtain a state-independent representation of our gate set, we shall instead work at the level of the covariance matrices, which completely characterize the Gaussian state. Hence one views the circuit as a matrix acting on the covariance matrix representation of the states in the usual manner; i.e., the evolution equation \eqref{eq:evo} becomes
\be
G_\mr{T}=UG_\mr{R}U^T~,\label{eq:evoGapx}
\ee
where the matrix elements of $G$ are given by
\be
G^{ab}=\bra{\psi}\xi^a\xi^b+\xi^b\xi^a\ket{\psi}~,\label{eq:Gbasis}
\ee
where $\xi^a\equiv\{x^1,p^1,\ldots x^N,p^N\}$ are the dimensionless phase-space operators for $N$ oscillators; cf.~\eqref{eq:evoG} in the main text.

Now, one has the freedom to choose the gate set generated by $M_I$ with which to build the circuit. Since the gate set ultimately determines the geometry, one will obtain different geodesics -- and hence different complexities -- for different choices. Provided the algebra that generates the set of gates is sufficient to produce the target state however, there are no rules for how to select it, and indeed several different choices have been analyzed in the literature. For example, \cite{Jefferson:2017sdb} chose a set of scaling and entangling gates that formed a representation of $\mathrm{GL}(N,\reals)$, while \cite{Chapman:2017rqy,Alves:2018qfv} worked instead with $\mathrm{SU}(1,1)$. In order to study the time-dependence of the TFD, \cite{TFD} chose a representation of $\Sp(2N,\reals)$, on the basis that this is the general group of transformations that preserves the canonical commutation relations. 

Given this freedom, we will be motivated by physical considerations in selecting our gates, namely, we wish to use the minimum set of gates sufficient to effect our particular class of target states. Since we shall work in the normal-mode basis, where the ground state factorizes as in \eqref{eq:psiProd}, it suffices to compute the complexity for a single mode (the generalization to $N$ oscillators is then straightforward, and is given in the main text). For the reference \eqref{eq:psiR} and target \eqref{eq:psiT} states under consideration, we have
\be
G_\mr{R}=\begin{pmatrix} \frac{1}{\omega_\mr{R}} & 0 \\ 0 & \omega_\mr{R} 
\end{pmatrix}~,\qquad
G_\mr{T}=\begin{pmatrix} \frac{1}{a} & -\frac{b}{a} \\ -\frac{b}{a} & \frac{a^2+b^2}{a} \end{pmatrix}~.\label{eq:Gs}
\ee
Hence we wish to select a set of gates that allows us to interpolate between these two states. This can be done with a subalgebra of $\frak{sp}(2,\reals)$. We shall denote the generators in the full algebra by
\be
W=\frac{i}{2}\lp xp+px\rp~,\,\,\,
V=\frac{i}{\sqrt{2}}x^2~,\,\,\,
Z=\frac{i}{\sqrt{2}}p^2~,\label{eq:full}
\ee
where $x,p$ are the dimensionless variables introduced below \eqref{eq:Hphys}. One can check that these generators satisfy the algebra
\ben
\comm{W}{V}=2V~,\qquad
\comm{W}{Z}=-2Z~,\qquad
\comm{V}{Z}=-2W~.\qquad
\een
Thus one sees that the $W$ and $V$ gates close to form a subalgebra. Hence, in contrast to \cite{TFD}, which analyzed circuits using the full algebra of $\frak{sp}(2,\reals)$, we shall restrict our circuits to the submanifold corresponding to the group elements
\be
X_W=e^{\eps W}~,\qquad
X_V=e^{\eps V}~,\label{eq:Xgates}
\ee
where $0<\eps\in\reals$. The corresponding matrix generators are then
\be
M_1=\begin{pmatrix} -1 & 0 \\ 0 & 1 \end{pmatrix}~,\qquad
M_2=\begin{pmatrix} 0 & 0 \\ \sqrt{2} & 0 \end{pmatrix}~,\label{eq:M}
\ee
where $M_1$ corresponds to $W$ and $M_2$ corresponds to $V$. Note that these are not orthonormal, but instead satisfy
\be
M_I^TM_J=2\,\delta_{IJ}~.\label{eq:norm}
\ee
The gates corresponding to \eqref{eq:Xgates} are then obtained by exponentiating these matrix elements:
\be
\hspace{-3 pt}Q_1=e^{\eps M_1}=\begin{pmatrix} e^{-\eps} & 0 \\ 0 & e^\eps \end{pmatrix},\,\,
Q_2=e^{\eps M_2}=\begin{pmatrix} 1 & 0 \\ \sqrt{2} \, \eps & 1 \end{pmatrix}.\label{eq:gates}
\ee
Let us pause briefly to remark on the scale $\varpi$, introduced in the dimensionless variables $x\!=\!\varpi X$, $p\!=\!P/\varpi$. We refer to this as the gate scale because, had we written the generators \eqref{eq:full} in terms of the dimensionful variables $X,P$, then both $V$ and $Z$ would contain factors of $\varpi^2$ in order to make the exponents in \eqref{eq:Xgates} dimensionless. Setting $\varpi=M$ is then a natural choice in order to avoid introducing an auxiliary scale into the problem, and we shall do so henceforth.

Now, to find the geometry in which the circuit \eqref{eq:Um} lives, we must find a suitable parametrization of the general group element with which to construct the metric. As shown in appendix B, a convenient choice is
\be
U=\frac{1}{\sqrt{z}}\begin{pmatrix} z & 0 \\ \frac{y}{\sqrt{2}} & 1 \end{pmatrix}~.\label{eq:Uxz}
\ee
The reason for this choice is that, upon isolating the components of the frame bundle in the usual manner \cite{Jefferson:2017sdb},
\be
\dd Y^I=\frac{1}{2}\tr{\dd UU^{-1}M_I^T}
\ee
(where the factor of $\frac{1}{2}$ is due to the normalization of the generators \eqref{eq:norm}), up to an unimportant overall normalization, the most general positive definite line element is
\be
\ba
\dd s^2&=g_{IJ}\dd Y^I\!\dd Y^J
=\frac{\dd z^2+A^2\dd y^2+2A\sin\sigma\dd y\dd z}{4z^2}~,
\ea
\ee
where we have taken
\be
g_{IJ}=\begin{pmatrix} 1 & -A\sin\sigma \\ -A\sin\sigma & A^2\end{pmatrix}~.
\label{eq:gPen}
\ee
In this expression, $A$ and $\sigma$ are penalty factors which account for different weighting of different gates. In contrast to metrics obtained in \cite{Jefferson:2017sdb,TFD}, their presence does not prevent us from having a closed form expression for geodesic length between two points $(z_0,y_0)$ and $(z_1,y_1)$, since we always deal with the metric on hyperbolic disc for which the distance function is known. However, for reasons which will become apparent below, we shall focus on the special case $\sigma\!=\!0$ and $A\!=\!2^{-1/2}$ for the remainder of this text. In this case, the distance function on the hyperbolic disc,
\be
\DD_{01}=\frac{1}{2}\ln\lp X_{01}+\sqrt{X_{01}^2-1}\rp,\label{eq:dist}
\ee
is evaluated with
\be
X_{01}\equiv\frac{2\lp z_1^2+z_0^2\rp+\lp y_1-y_0\rp^2}{4 z_1z_0}~.\label{eq:X01}
\ee
Now, the optimum circuit is the minimum geodesic on $\mathbb{H}^2$ that connects the reference $G_\mr{R}$ and target $G_\mr{T}$ states, and the complexity of the latter is given by the length of this circuit, \eqref{eq:dist}. It therefore remains simply to express the initial and final coordinates, $(z_0,y_0)$ and $(z_1,y_1)$, in terms of the physical parameters of the problem at hand, i.e., the frequencies of the reference and target states. 

However, one important caveat is in order: the generalization of this prescription to $N$ oscillators relies on the fact that for our quench solution, both the reference and target states remain block diagonal. When considering purifications however, we begin with a dense covariance matrix representing the target state, and must first bring it to block-diagonal form prior to computing the complexity. This can be done by defining a squeezing operator
\be
S=\mathrm{diag}\{\sqrt{\omega_1},\,\frac{1}{\sqrt{\omega_1}},\,\ldots,\,\sqrt{\omega_N},\,\frac{1}{\sqrt{\omega_N}}\}~,
\ee
where $\omega_1$ through $\omega_N$ are the reference state frequencies for each of the $N$ oscillators. The definition of $S$ is such that its action on the reference state produces the identity matrix,
\be
SG_\mr{R}S^T=\mathbbm{1}~.
\ee
Applying this operator to the target state then allows us to block-diagonalize the latter without introducing off-diagonal terms in the reference state. Therefore in order to compute complexity for such states, we will evaluate the distance \eqref{eq:X01} between the identity and the squeezed target state
\be
\tilde G_\mr{T}\equiv SG_\mr{T}S^T=\begin{pmatrix} \frac{\omega_\mr{R}}{a} & -\frac{b}{a} \\ -\frac{b}{a} & \frac{a^2+b^2}{\omega_\mr{R} a} \end{pmatrix}~.
\ee

To proceed, we impose boundary conditions on the circuit $U$ \eqref{eq:Uxz}, where $z(s), y(s)$ are functions of the path parameter $s\in[0,1]$ (that is, \eqref{eq:Uxz} is merely the coordinate representation of \eqref{eq:Um}). At $s\!=\!0$, the circuit has not produced any change in the reference state, and hence the initial condition is
\be
U(0)G_\mr{R}U(0)^T=G_\mr{R}\;\implies\;U(0)=\mathbbm{1}~,
\ee
which holds for $G_\mr{R}$ in \eqref{eq:Gs} for arbitrary $\omega_\mr{R}$. Conversely, at $s\!=\!1$ the circuit should satisfy \eqref{eq:evoGapx}, with $G_\mr{R}\!\rightarrow\!\mathbbm{1}$ and $G_\mr{T}\!\rightarrow\!\tilde G_\mr{T}$, which enables one to solve for the final coordinates in terms of the physical frequencies; one finds:
\ben
\lp z_0,~y_0\rp=\,\lp1,0\rp~,\qquad
\lp z_1,~y_1\rp=\lp\frac{\omega_\mr{R}}{a},-\frac{\sqrt{2}\,b}{a}\rp~.
\een
Substituting these into the expression for the geodesic length \eqref{eq:dist}, we obtain \eqref{eq:C}. Due to the ability to simultaneously diagonalize $G_\mr{R}=\mathbbm{1}$ and $\tilde G_\mr{T}$, the result for $N$ oscillators is simply $N$ copies of the hyperbolic disc.

As alluded above, one can also explicitly solve the geometry in the presence of the general penalty factors in \eqref{eq:gPen}. These were originally introduced in \cite{Jefferson:2017sdb} in order to impose a notion of locality on the circuit, without which the gates are non-local; the interested reader is referred to the discussion therein for more details. In this case, the result is rather unwieldy and unilluminating, so we refrain from writing it out here. However, it is worth noting that the scaling behaviours observed in the main text are insensitive to the choice of penalty factors. The reason is two-fold: first, it is a general fact that one can absorb the penalty factor $A$ by rescaling the imaginary component of the target state $b\rightarrow b/A$. Second, it turns out that when specifying to the quench solution considered in the main text, this imaginary component vanishes for the full parameter range. Therefore the complexity only depends on the penalty factors through $\sin\sigma$, and one can furthermore show that the unique range of this parameter is $\sigma\in[0,\pi/2)$ (the complexity diverges at precisely $\sigma\!=\!\pi/2$). For values of $\sigma$ within this range, the effect is merely a constant shift in Fig.~\ref{fig:QC-scalings}, and a slightly more gradual transition between the linear and logarithmic $\tfrac{1}{4}\ln\delta t$ regimes. The value of the coefficient in the latter may be obtained by numerical fitting, and is therefore sensitive to the width of this transition zone, but remains largely unchanged.

\section{B. Geodesics on $\mathbb{H}^2$}\label{appx:geometry}

In this appendix, we show that the geometry corresponding to the use of the two gates \eqref{eq:Xgates} can be derived as an embedding in AdS, which enables us to readily obtain the geodesic distance \eqref{eq:dist}. We begin by finding a suitable parametrization for the general group element, which will represent the circuit $U(s)$ \eqref{eq:Um}. Of course, the most na\"ive way to express the generic group element generated by \eqref{eq:M} is
\be
U=\exp\left\{\mu M_1+\nu M_2\right\}
=\begin{pmatrix}
e^{-\mu} & 0 \\ \frac{\sqrt{2}\nu}{\mu}\sinh\mu\; & e^\mu
\end{pmatrix}
\ee
where $\mu,\nu\in\reals$. However, by making the change of variables
\be
\mu=-\frac{1}{2}\ln z~,\qquad
\nu=\frac{1}{2}\frac{y}{z-1}\ln z~,
\ee
this becomes
\ben
U=\frac{1}{\sqrt{z}}\begin{pmatrix} z & 0 \\ \frac{y}{\sqrt{2}} & 1 \end{pmatrix}~.
\een
which is \eqref{eq:Uxz}. As mentioned above, the reason for this choice is that the resulting metric becomes that of the hyperbolic plane. In particular, choosing $\sigma=0$, $A=2^{-1/2}$ yields
\be
\dd s^2=\ell^2\frac{2 \dd z^2+\dd y^2}{2z^2}~,\label{eq:ds}
\ee
where $\ell\equiv1/2$. One can reproduce $\mathbb{H}^2$ via the standard embedding of the pseudosphere in $\mathbb{R}^{d,1}$:
\be
\lp X^0\rp^2-\lp X^1\rp^2+\lp X^2\rp^2=-\ell^2~,\label{eq:embed}
\ee
where the metric of the embedding space is
\be
\dd s^2=-\eta_{MN}\dd X^M\!\dd X^N~,\quad
\eta_{MN}=\mr{diag}\{-1,1,-1\}~.\label{eq:dsMN}
\ee
One can check that the constraint \eqref{eq:embed} allows the following choice of parameters:
\ben
X^0=\frac{\ell y}{\sqrt{2}z}~,\;\;
X^1=\frac{\ell^2+z^2+y^2/2}{2z}~,\;\;
X^2=\frac{-\ell^2+z^2+y^2/2}{2z}~,
\een
upon which \eqref{eq:dsMN} reproduces \eqref{eq:ds}. Then, for points $(z_0,y_0)$ and $(z_1,y_1)$,
\be
\ell^2X_{01}\equiv\eta_{MN}X^M(z_0,y_0)X^N(z_1,y_1)~,
\ee
yields precisely \eqref{eq:X01}, with the geodesic distance given by \eqref{eq:dist}. 

\section{C. Complexity of purification}\label{appx:cpurification}
In this appendix, we explain how to compute the complexity of purification for the two-oscillator system considered in the main text, i.e., to associate a complexity to the subsystem consisting of a single oscillator. The position-space wavefunction for the total system is
\be
\psi\lp x_1,x_2\rp=\mathcal{N}\exp\left(-\frac{\omega_{1}}{2}x^2_1-\frac{\omega_{2}}{2}x^2_2-\omega_3x_1x_2\right)~,\label{eq:totgauss}
\ee
where $\omega_{i}\!\equiv\! a_{i}\!+\!ib_{i}$ with $a_i,b_i\!\in\!\mathbb{R}$, and the normalization is $\mathcal{N}\!=\! (a_{1}a_{2}\!-\!a_{3}^2)^{1/4}/\sqrt{\pi}$, cf.~\eqref{eq:psiT}. Note that we must have $a_i>0$ and $a_1a_2>a_3^2$ in order for the wavefunction to be well-defined. The subsystem corresponding to the first oscillator is described by the reduced density matrix
\be
\rho_{1}(x_{1},y_{1})=\!\sqrt{\frac{\pi}{a_2}}\mathcal{N}^2\,e^{-\frac{\alpha_1}{2}\left(x^2_1+y^2_1\right)+\beta x_1 y_1}\,e^{-\frac{i\alpha_2}{2}\left(x^2_1-y^2_1\right)}~,\label{eq:RDM1}
\ee
where we have defined
\ben
\alpha_{1}\equiv a_{1} - \frac{a_{3}^{2}-b_{3}^{2}}{2a_{2}}~,\quad
\alpha_{2}\equiv -b_{1}+ \frac{a_{3}b_{3}}{a_{2}}~,\quad
\beta\equiv\frac{a_{3}^{2}+b_{3}^{2}}{2a_{2}}~.
\een
Thus, of the original six (real) parameters on which the total system \eqref{eq:totgauss} depends, three are fixed by specifying data for the subsystem. Our strategy is then to minimize the complexity over the remaining three parameters. That is, we purify the subsystem with a second oscillator, so that the total wavefunction can be written in the form \eqref{eq:totgauss}, and the parameters of this purification are those which effect the minimum complexity.

To rephrase this in the language of covariance matrices employed in the main text: the two-oscillator system \eqref{eq:totgauss} is described by a four-by-four covariance matrix. In the basis $\xi^a=\{x^1,p^1,x^2,p^2\}$ (cf.~\eqref{eq:Gbasis}), this is naturally organized into diagonal two-by-two blocks that describe the subsystems (i.e., the individual oscillators), and off-diagonal blocks which describe the entanglement between them. It is straightforward to show that specifying the three parameters $\alpha_1,\alpha_2,\beta$ in the reduced density matrix \eqref{eq:RDM1} completely fixes the first diagonal block corresponding to the $x_1$ oscillator; explicitly,
\be
G_\mr{1}=\begin{pmatrix} \frac{1}{\alpha_{1}-\beta} & \frac{\alpha_{2}}{\alpha_{1}-\beta} \\ \frac{\alpha_{2}}{\alpha_{1}-\beta} & \frac{\alpha_{1}^{2}+\alpha_{2}^{2}-\beta^{2}}{\alpha_{1}-\beta} 
\end{pmatrix}~.
\label{eq:G1}
\ee
Note that this corresponds to fixing the expectation values $\langle x_{1}^2 \rangle$, $\langle p_1^2\rangle$, and $\langle x_1p_1+p_1x_1\rangle$; see the discussion below Fig.~\ref{fig:purification}.

We then apply our formula for the complexity \eqref{eq:Cfield} with $N\!=\!2$,
\be
\CC=\frac{1}{2}\sqrt{ \ln\lp\chi_{1}+\sqrt{\chi_{1}^2-1}\rp^2+\ln\lp\chi_{2}+\sqrt{\chi_{2}^2-1}\rp^2},
\label{eq:C2osc}
\ee
where $\chi_i\!=\!\frac{1}{2}\mr{tr}\,\tilde G_i$, and $\tilde G_i\!=\!SG_iS^T$ is the squeezed covariance matrix, cf.~\eqref{eq:C}. This expression depends on six real parameters, three of which are fixed in terms of $\alpha_1,\alpha_2,\beta$. To simplify the numerical minimization procedure over the remaining three parameters, it is convenient to perform a change of variables to spherical coordinates, so that two of them become compact; the range of the remaining non-compact (radial) parameter can then be constrained by experimentation. 

After repeating this process for the second oscillator (that is, fixing the parameters in the reduced density matrix $\rho_2$, which are \emph{a priori} independent), we can compute the subsystem complexities $\CC_A$ and $\CC_{\bar A}$ used in verifying the superadditivity property \eqref{eq:super}. It would of course be interesting to test this conjecture for varying subsystem sizes, rather than the fixed equipartition forced upon us by the two-oscillator case. Unfortunately, it is computationally challenging to extend this procedure to $N\!>\!2$, but we hope to return to this question in future work.

Let us briefly contrast the complexity of purification with the information accessed by entanglement. Consider the case in which we know the reduced density matrices for both subsystems, which for simplicity we may fix symmetrically, i.e., $\langle x_{1}^2 \rangle = \langle x_{2}^2 \rangle$, $\langle p_{1}^2 \rangle = \langle p_{2}^2 \rangle$ and $\langle x_{1} \, p_{1} \rangle = \langle x_{2} \, p_{2} \rangle$. This leaves a one-parameter freedom in the full covariance matrix, which we can choose to be $\langle x_{1} \, p_{2} \rangle = \langle x_{2} \, p_{1} \rangle$. Obviously, the entanglement entropy for either subsystem will be insensitive to the information this correlator contains about the total state. In contrast, the complexity varies as a function of this parameter, and therefore provides a complementary probe to entanglement for a fixed bipartition. We stress however that this dependence is not necessarily monotonic, which prevents us from directly identifying complexity as an order parameter for the wavefunction. It would certainly be interesting to investigate the relative merits of entanglement and complexity in this regard, and the reader is referred to the upcomming work \cite{notyet} for a more thorough treatment of entanglement in this context.
\\

\section{D. Scalings and choice of norm}\label{appx:normscale}

Here we comment briefly on how the scaling behaviours of the critical complexity emerge in the field theory in different norms. As shown in Fig.~\ref{fig:QC-scalings}, the zero-mode complexity exhibits a linear growth with $\delta t$ which smoothly transitions to the $\log\!\delta t$ KZ behaviour at $\delta t^{-1}\!\sim\!\omega_0$. As we move towards the UV, both the exact solution as well as the KZ approximation for non-zero modes (see Fig.~\ref{fig:QC-mode}) saturate to smaller values of complexity at lower values of $\delta t$. Thus when adding the modes using the $L_p$-norm (with $p\!\geq\!2$), the field theory critical complexity is dominated by the zero mode and hence inherits its scaling behaviour. If we instead use the $L_1$-norm, the mode-by-mode saturation values are no longer sufficiently suppressed for the field theory to simply inherit the scaling behaviour of the zero mode, and thus it is unclear whether universal scalings can be extracted in this case. However, previous studies \cite{Jefferson:2017sdb,Chapman:2017rqy} found that $L_1$ exhibits better agreement with holography than $L_2$. Hence it would be very interesting to explore critical quenches holographically (non-critical quenches have been studied in, e.g., \cite{Moosa:2017yvt,Chapman:2018dem,Chapman:2018lsv,Ageev:2018nye}). 

Meanwhile in the case of entanglement entropy studied in~\cite{Caputa:2017ixa}, the authors observed only a narrow range in which the fast scaling is applicable with quadratic dependence on the quench rate $\delta t$. We believe that the range of fast scalings in~\cite{Caputa:2017ixa} depends on the subsystem size. The analogue of this additional dimensionful parameter in the case of complexity is the reference state scale, which indeed controls the range over which we observe fast scalings. Additionally, we note that in this work we consider field theories on a circle; in the decompactification limit, we expect the zero mode to give a subdominant contribution to the dynamics, as observed recently in \cite{TFD}.

\end{document}